\documentclass[a4pape,citesort]{PoS}
\usepackage{caption}
\usepackage{subcaption}
\usepackage{graphicx}
\usepackage{nicefrac}
\usepackage{citesort}
\usepackage{wrapfig}

\usepackage{caption,setspace}
\usepackage{lipsum,lineno}
\usepackage[utf8]{inputenc}
\usepackage[none]{hyphenat}
\makeatletter 
\renewcommand\speaker[1]{\if@speaker\global\@dblspeaktrue\fi
			\global\@speakertrue
			\global\setbox\@firstaubox
			\hbox{{\let\thanks\@gobble
				\let\footnote\@gobble\small
				\rm  The xFitter Developers' Team}}%
			#1\thanks{Speaker.}\
			}%
\makeatother

\title{xFitter 2.0.0: Heavy quark matching scales: \hspace{\textwidth} 
{\it Unifying the FFNS and VFNS}
}

\ShortTitle{xFitter 2.0.0: Heavy Quark Matching Scales }

\def\thanksref#1{\rlap,${}^{#1}$}
\def\inst#1{\hangafter=1\hangindent=15pt\relax ${}^{#1}$}

\author{
The xFitter Developers' Team:\thanks{%
We acknowledge the hospitality of CERN, DESY, and Fermilab where a
portion of this work was performed.
This work was also partially supported by the U.S.\ Department of
Energy under Grant No.\ DE-SC0010129. 
We are grateful to the DESY IT department for their support of the
xFitter developers.
}
\quad
V.~Bertone\thanksref{a,b}  \ 
D. Britzger\thanksref{c} \ 
S.~Camarda\thanksref{d}   \ 
A.~Cooper-Sarkar\thanksref{e} \ 
A.~Geiser\thanksref{c} \ 
F.~Giuli\thanksref{e} \ 
A.~Glazov\thanksref{c} \ 
E.~Godat\thanksref{f} \ 
A.~Kusina\thanksref{g,h} \ 
A.~Luszczak\thanksref{i}   \ 
F.~Lyonnet\thanksref{f} \ 
F.~Olness\thanksref{f}\speaker{} \ 
R.~Pla\v cakyt\.e\thanksref{j} \ 
V.~Radescu\thanksref{c,d}  \ 
I.~Schienbein\thanksref{g} \ 
and \
O.~Zenaiev\thanksref{c} \ 
\\
\inst{a} Department of Physics and Astronomy,  VU University, NL-1081 HV Amsterdam, The~Netherlands   \\
\inst{b} Nikhef Theory Group Science Park 105, 1098 XG Amsterdam, The Netherlands   \\
\inst{c} DESY Hamburg, Notkestra{\ss}e 85, D-22609, Hamburg, Germany    \\
\inst{d} CERN, CH-1211 Geneva 23, Switzerland    \\
\inst{e} University of Oxford,1 Keble Road, Oxford OX1 3NP, United Kingdom    \\
\inst{f} SMU Physics, Box 0175 Dallas, TX  75275-0175, United States of America    \\
\inst{g} Laboratoire de Physique Subatomique et de Cosmologie, Universit\'e Grenoble Alpes, \\
     \null \qquad    CNRS/IN2P3,    53 avenue des Martyrs, 38026 Grenoble, France   \\
\inst{h} Institute of Nuclear Physics, Polish Academy of Sciences,  ul. Radzikowskiego 152, 31-342 Cracow, Poland    \\
\inst{i}  T.Kosciuszko Cracow University of Technology, 30-084 Cracow, Poland   \\
\inst{j} Institut f\"ur Theoretische Physik, Universit\"at Hamburg, Luruper Chaussee 149, D--22761 Hamburg, Germany    
}
\abstract{%
xFitter~\cite{Alekhin:2014irh}  is an open-source package that provides a framework for the
determination of the parton distribution functions (PDFs) of the
proton  for many different kinds of analyses in Quantum
Chromodynamics (QCD). 
It incorporates experimental data from a wide range of experiments 
including fixed-target, Tevatron, HERA, and LHC. 
xFitter version 2.0.0 has recently been released, and offers an expanded set of tools and options. 
The new xFitter~2.0.0 program links to the
APFEL code~\cite{Bertone:2013vaa} which has implemented generalized
matching conditions that enable the switch from $N_F$ to
$N_{F}+1$ active flavors at an arbitrary matching scale $\mu_m$.
This enables us to  generalize the transition between
a FFNS and a VFNS and essentially vary continuously between 
the two schemes; in this sense the matching scale $\mu_m$ 
allows us to unify the FFNS and VFNS in a common framework~\cite{Bertone:2017ehk}.
This paper provides a brief overview of xFitter 
with emphasis of these new  features. 
}

\FullConference{XXVI International Workshop on Deep-Inelastic Scattering and Related Subjects (DIS2018)\\
		16-20 April 2018\\
		Kobe, Japan}

\begin{document}

\def\figone{
\begin{figure}[th]
\null
\vspace{-0.3cm}
  \centering
    \includegraphics[width=0.90\textwidth]{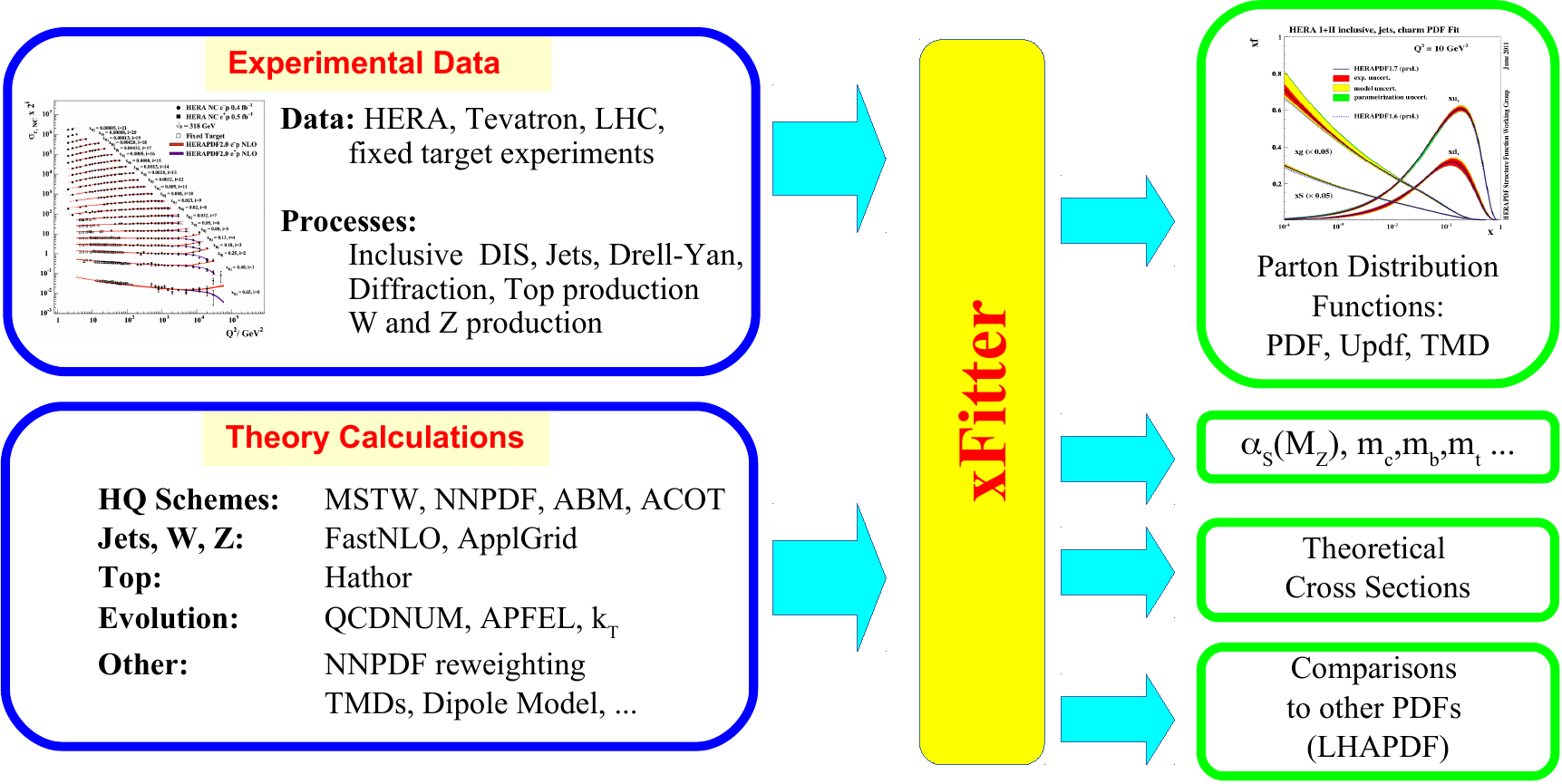}
    \caption{Schematic  of the modular structure of  xFitter  
illustrating the components and capabilities of the program.  \label{fig:flow}
}
\vspace{-0.8cm}
\end{figure}
}

\def\figfrog{
\begin{wrapfigure}{R}{2.9cm} 
\centering{} 
\vspace{-40pt}
\includegraphics[width=2.9cm]{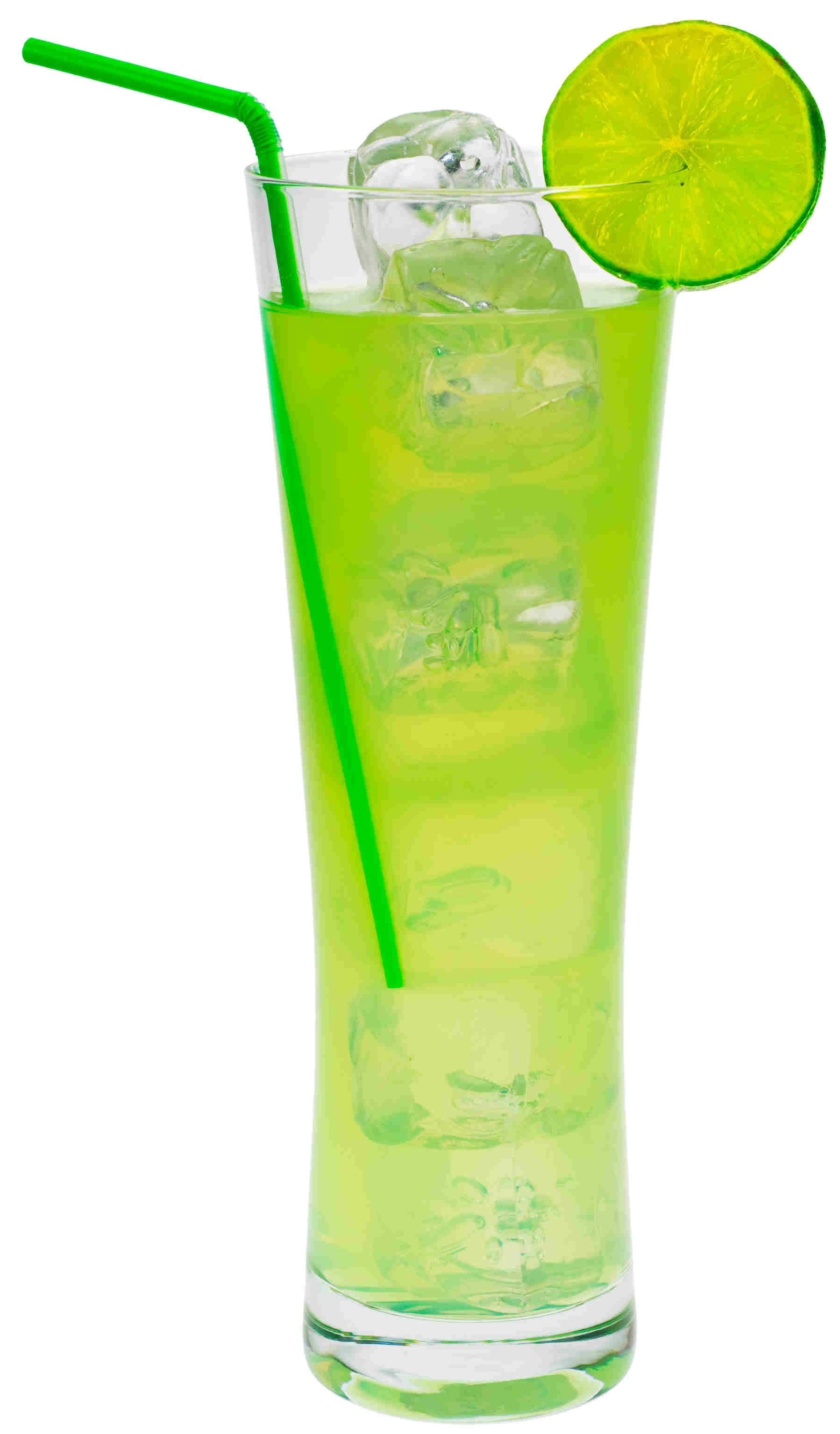}
\vspace{-20pt}
\caption*{
\linespread{0.9}\selectfont{}
\centering{} 
\null
\centerline{xFitter~2.0.0 \quad} 
\centerline{\it (Frozen~Frog)}
{\tiny iStock.com/Enjoylife2}
\vspace{-10pt}
}
\label{fig:one}
\end{wrapfigure}
}

\def\figvfnsi{
\begin{figure*}[t]
\centering{}
\includegraphics[width=0.45\textwidth]{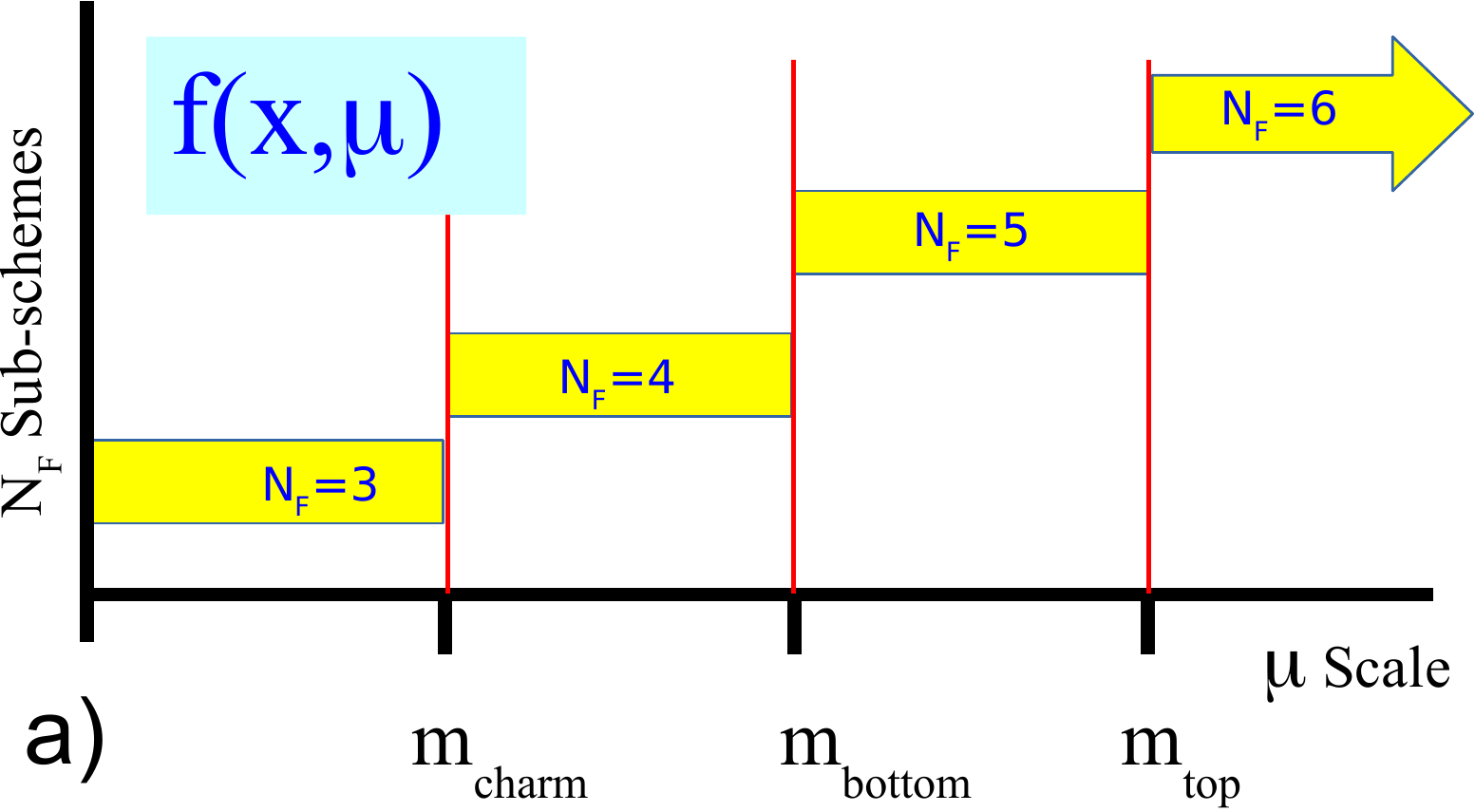}
\hfil
\includegraphics[width=0.45\textwidth]{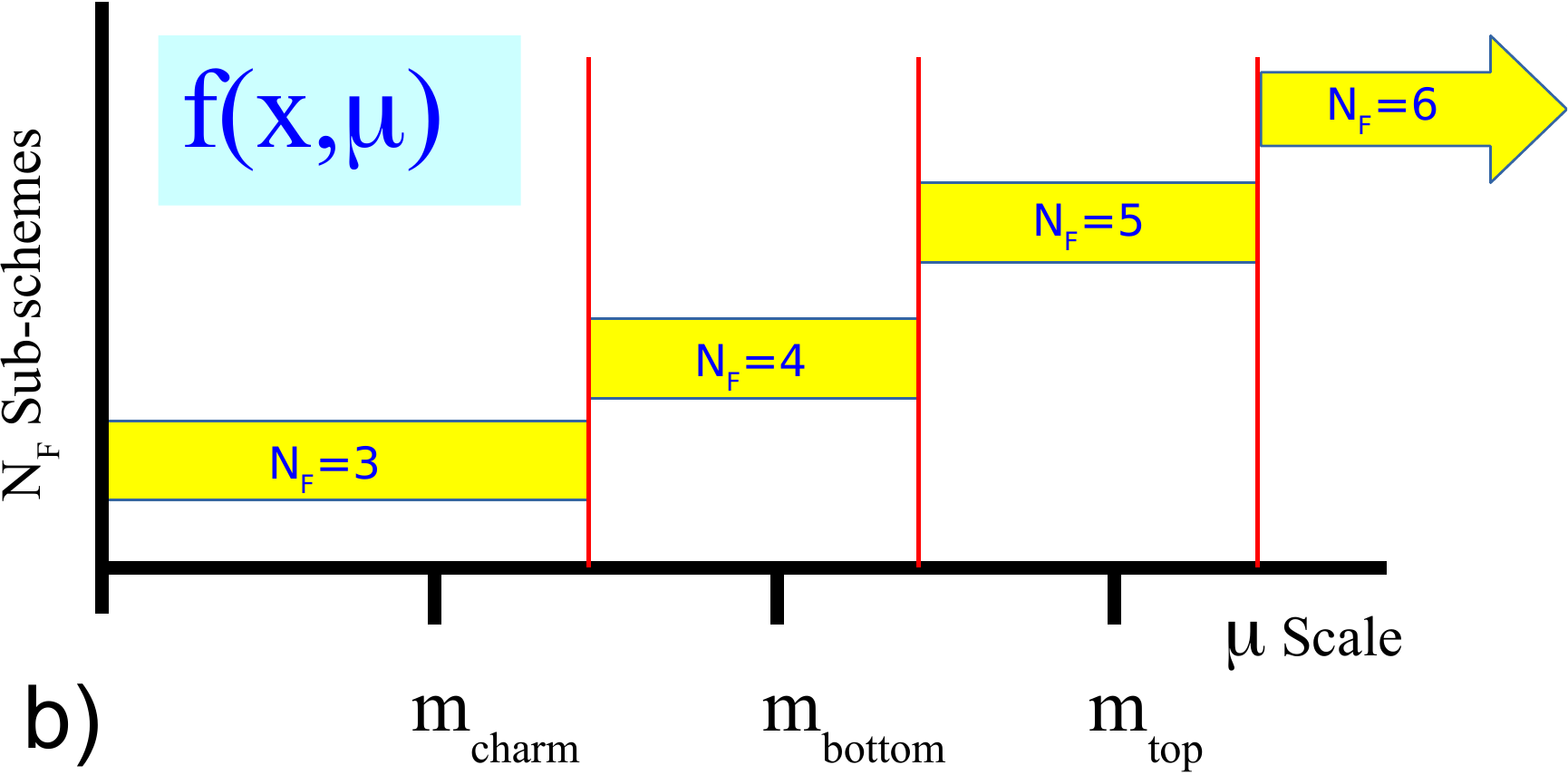}
\caption{An illustration of the separate $N_{F}$ renormalization sub-schemes
which define a VFNS. 
Historically, the matching scales $\mu_{m}$ 
were chosen to be exactly the mass values $m_{c,b,t}$ as in Fig.-a.
\quad 
Fig.-b is a generalized case where the   $\mu_{m}$ scales
are chosen to be different from the mass values. \label{fig:vfns}
}
\vspace{-0.5cm}
\end{figure*}
} %

\def\figBmatch{
\begin{figure*}[t]
\centering{}
\includegraphics[width=0.45\textwidth,angle=0]{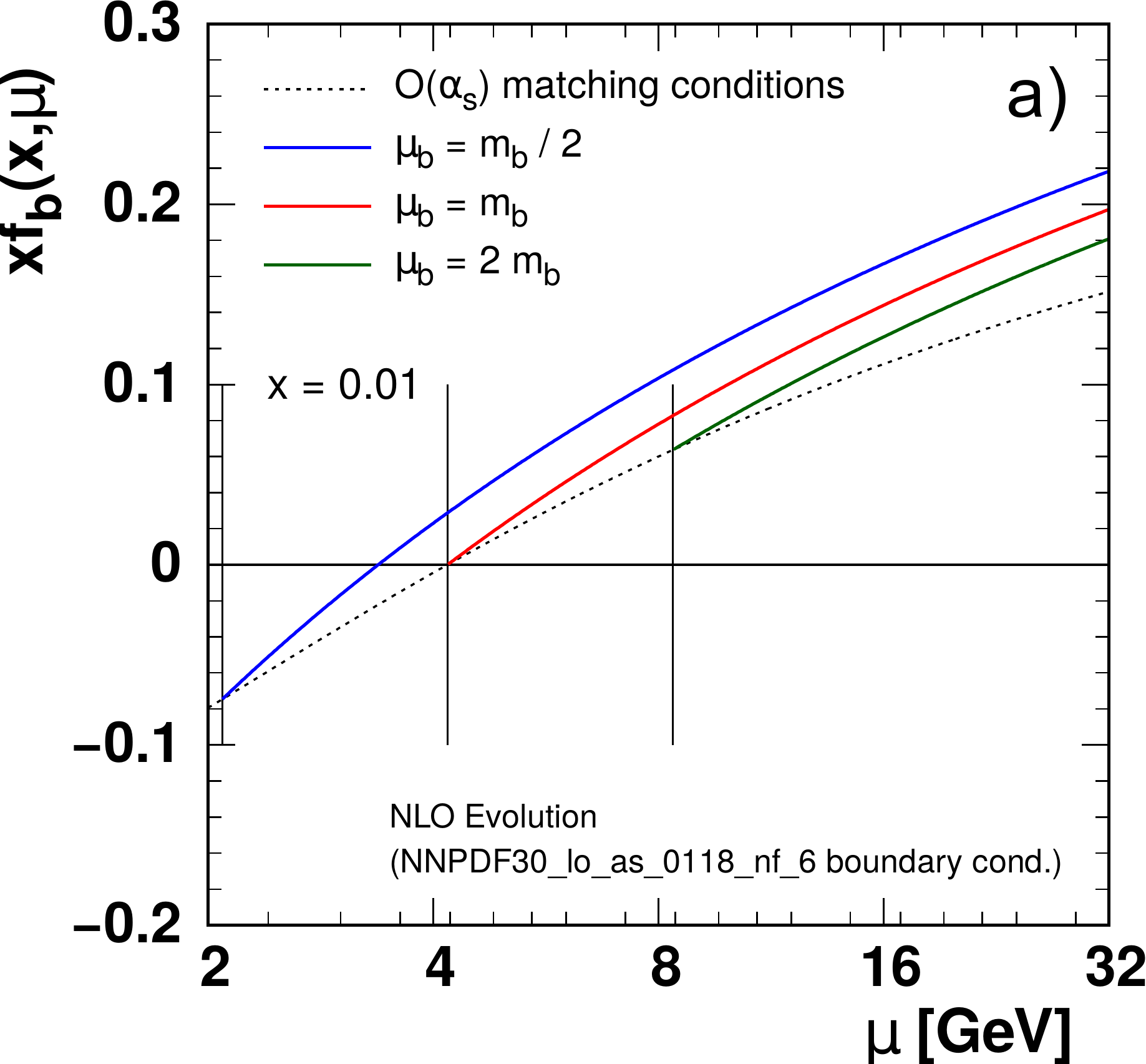}
\hfil
\includegraphics[width=0.45\textwidth,angle=0]{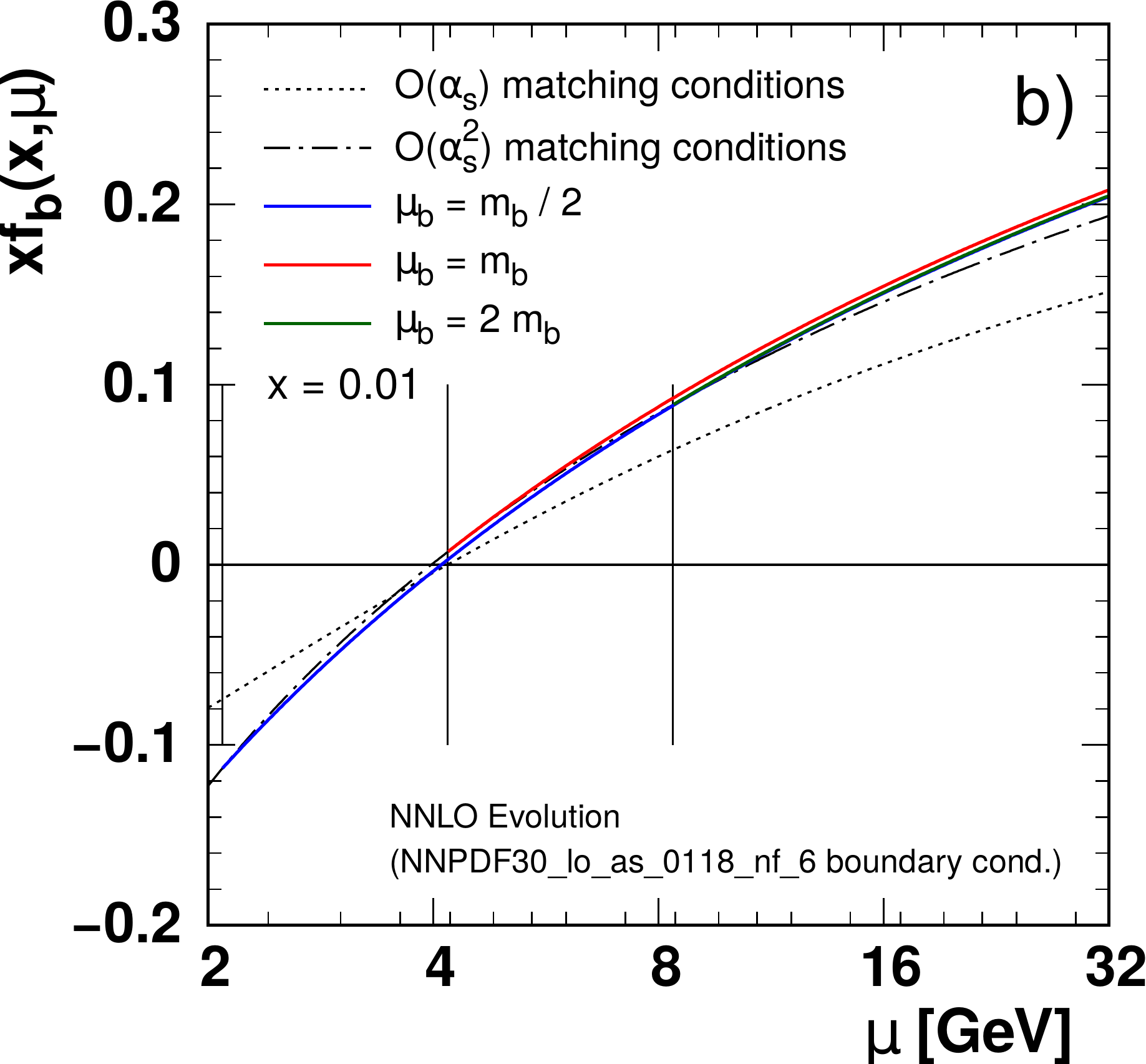}
\caption{We display the b-quark PDF $x\, f_b^{(5)}(x,\mu)$ for different choices
  of the matching scales 
$\mu_{m}=\{m_b/2,m_b,2m_b \}$ (indicated by the vertical lines)
computed at 
NLO (Fig.-a)
and NNLO (Fig.-b).
\label{fig:bMatch}
}
\vspace{-0.5cm}
\end{figure*}
} %

\def\figFbottom{
\begin{figure*}[t]
\centering{}
\includegraphics[width=0.45\textwidth]{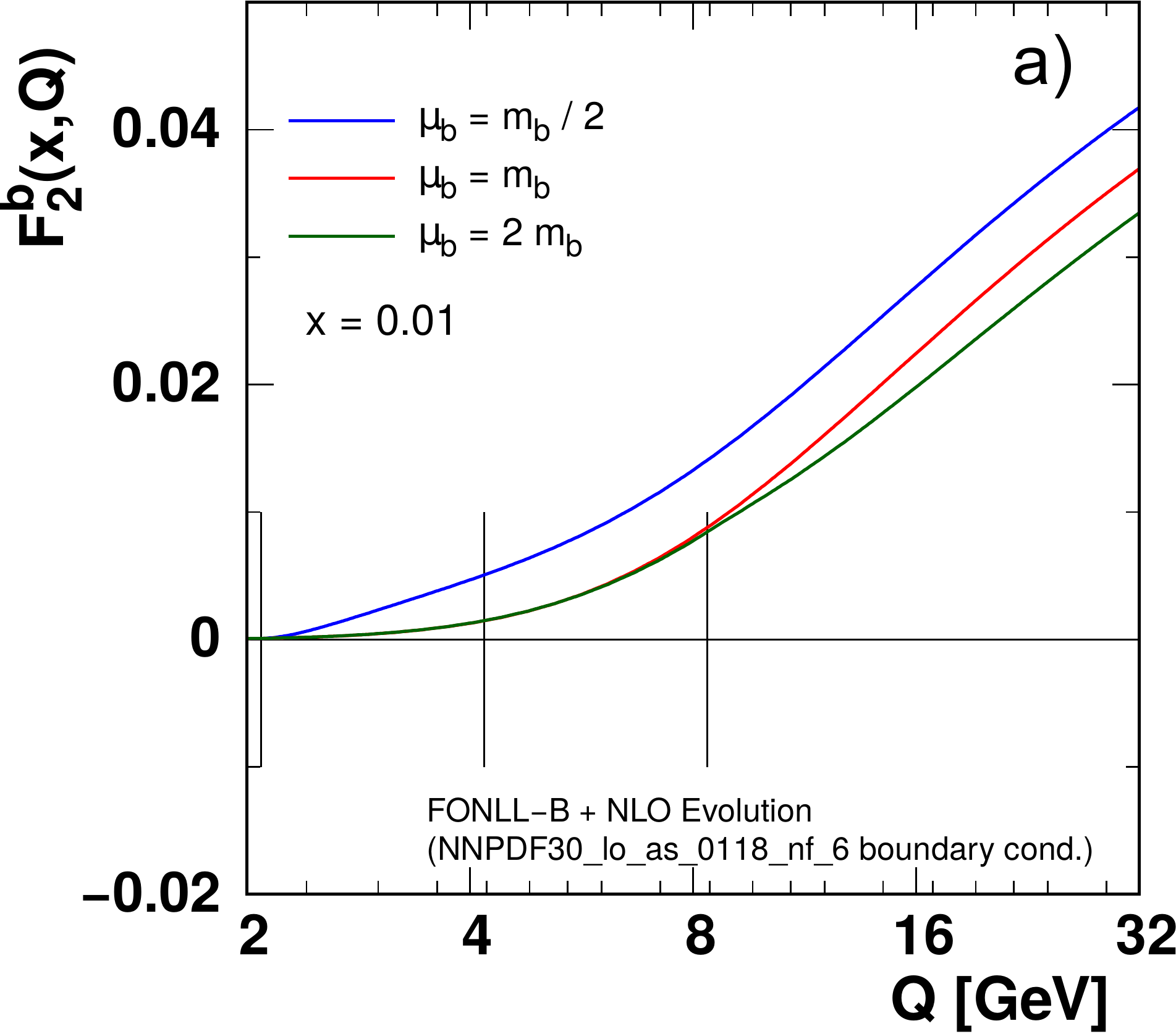}
\hfil
\includegraphics[width=0.45\textwidth]{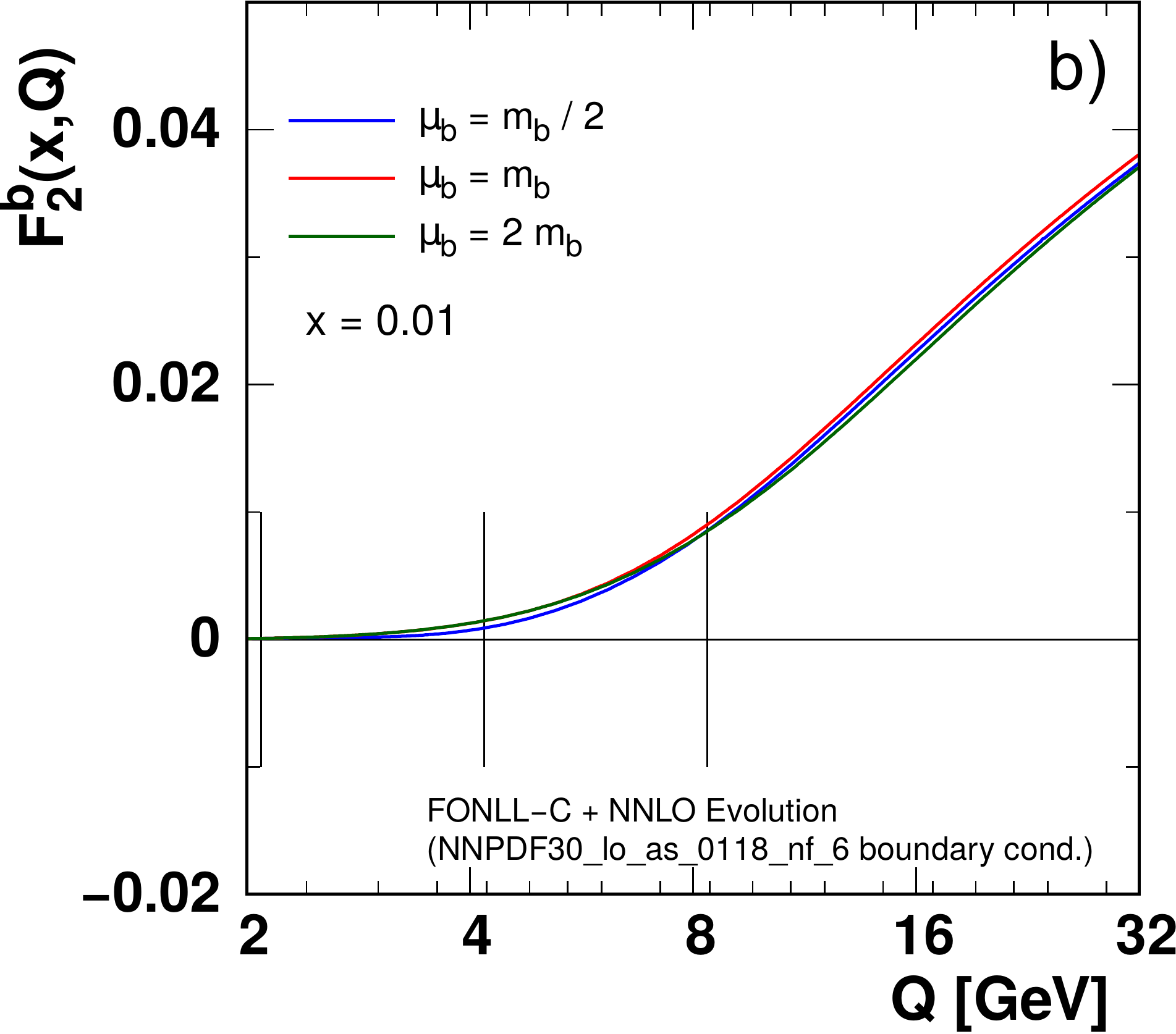}
\caption{We display$F_2^b(x,Q)$  for different choices
  of the matching scales 
$\mu_{m}=\{m_b/2,m_b,2m_b \}$ (indicated by the vertical lines)
computed at 
NLO (Fig.-a)
and NNLO (Fig.-b).
Here, we have chosen $\mu=Q$.
For details on the FONNL calculation see Ref.~\cite{Forte:2010ta}.
\label{fig:fbottom}
}
\vspace{-0.5cm}
\end{figure*}
} %

%
\def\figNfPDF{
\begin{figure*}[t]
\centering{}
\includegraphics[width=0.45\textwidth]{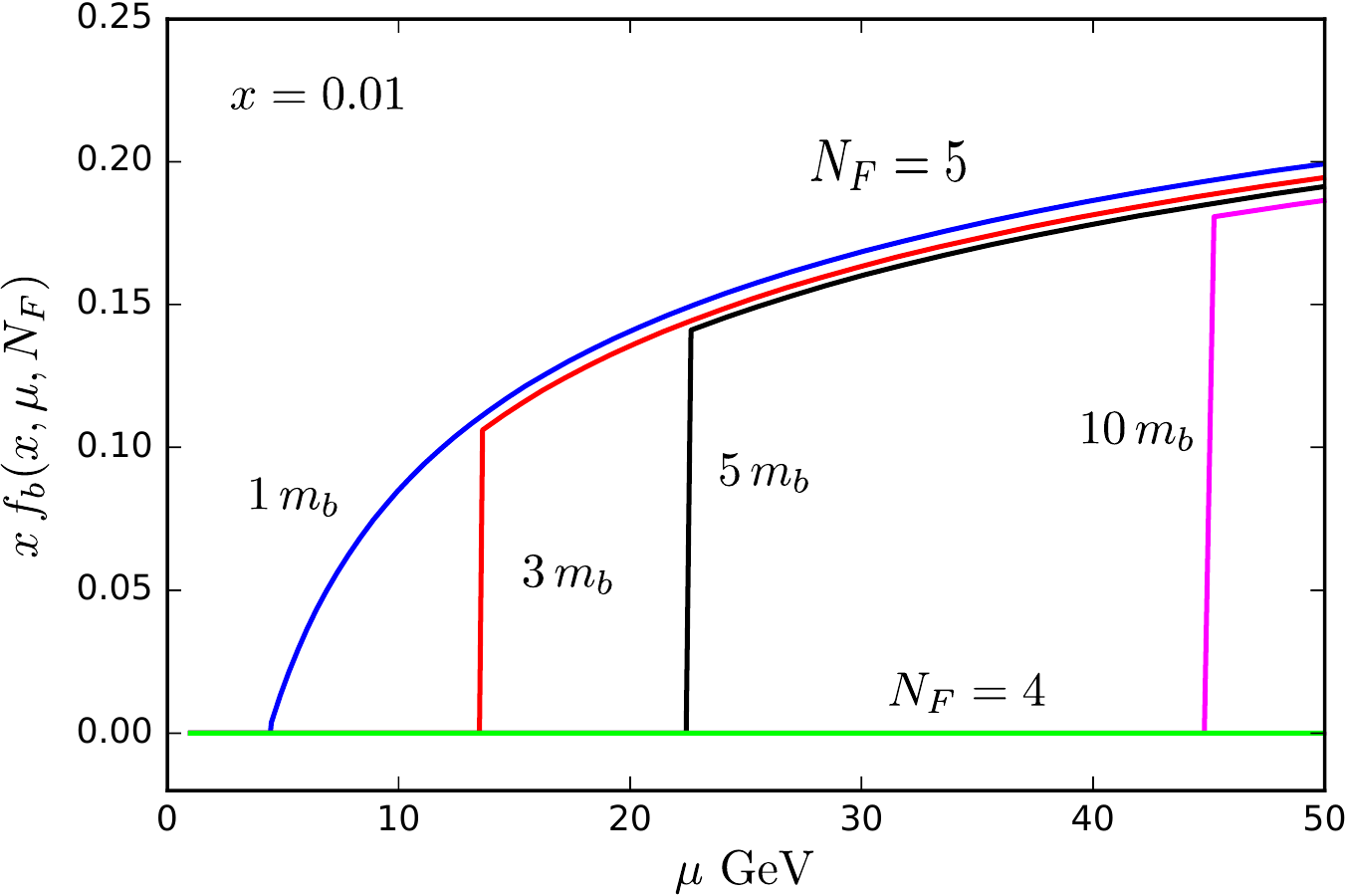}
\hfil
\includegraphics[width=0.45\textwidth]{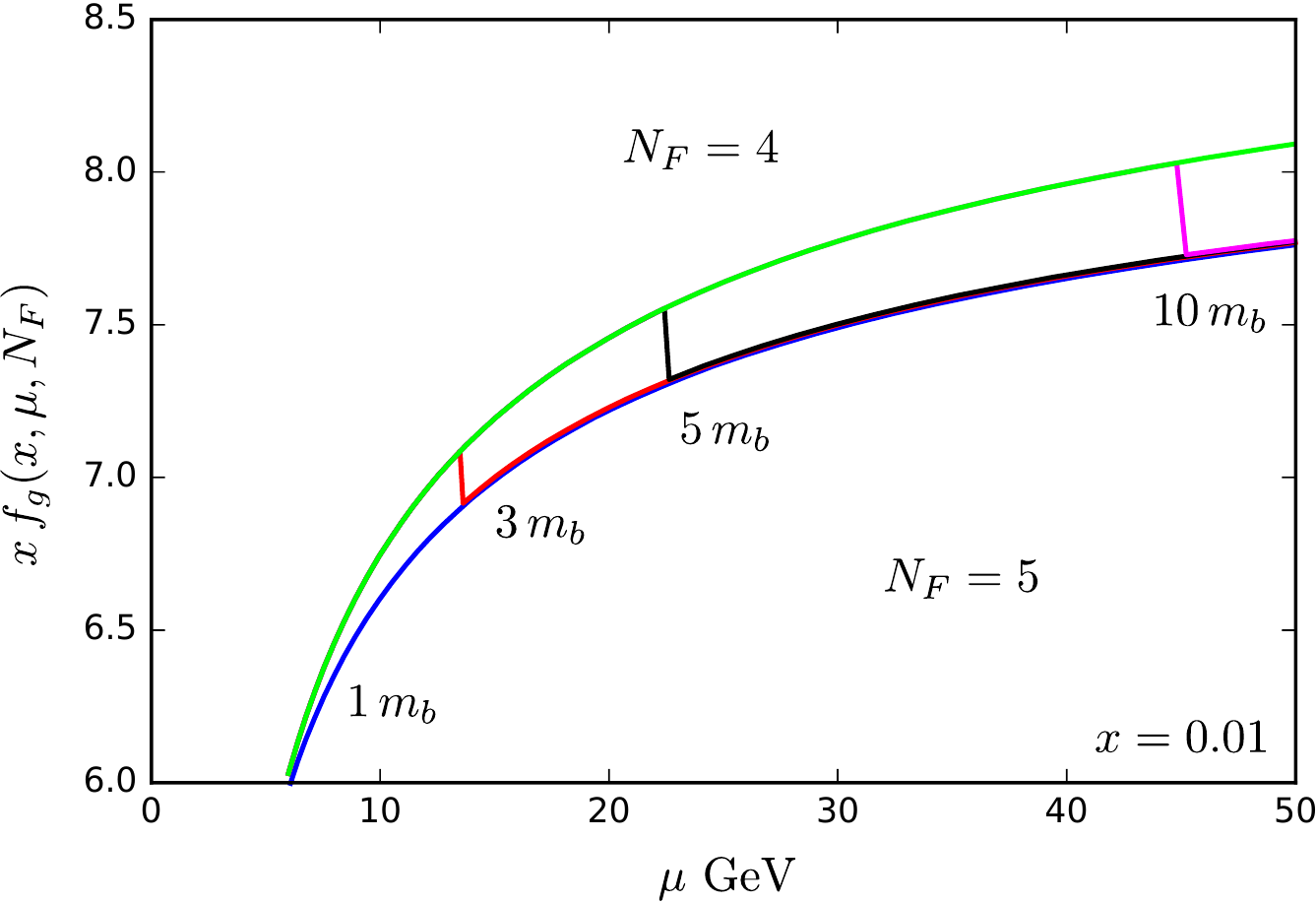}
\caption{$N_F$-dependent PDFs $x \, f_i(x,\mu,N_F)$ for the bottom quark (left) and gluon (right)
with  variable matching scales for 
$\mu_b=\{1,3,5,10, \infty\}\times m_b$ \{blue, red, black, magenta, green\}
with $x=0.01$ as a function of $\mu$ in GeV.  
The vertical lines in the plots show the transition from the $N_F=4$
to $N_F=5$ flavor scheme. 
\label{fig:pdfVsQ}
}
\end{figure*}
} %
%
\def\figChiScaledi{
\begin{figure*}[t]
\centering{}
\includegraphics[width=0.45\textwidth]{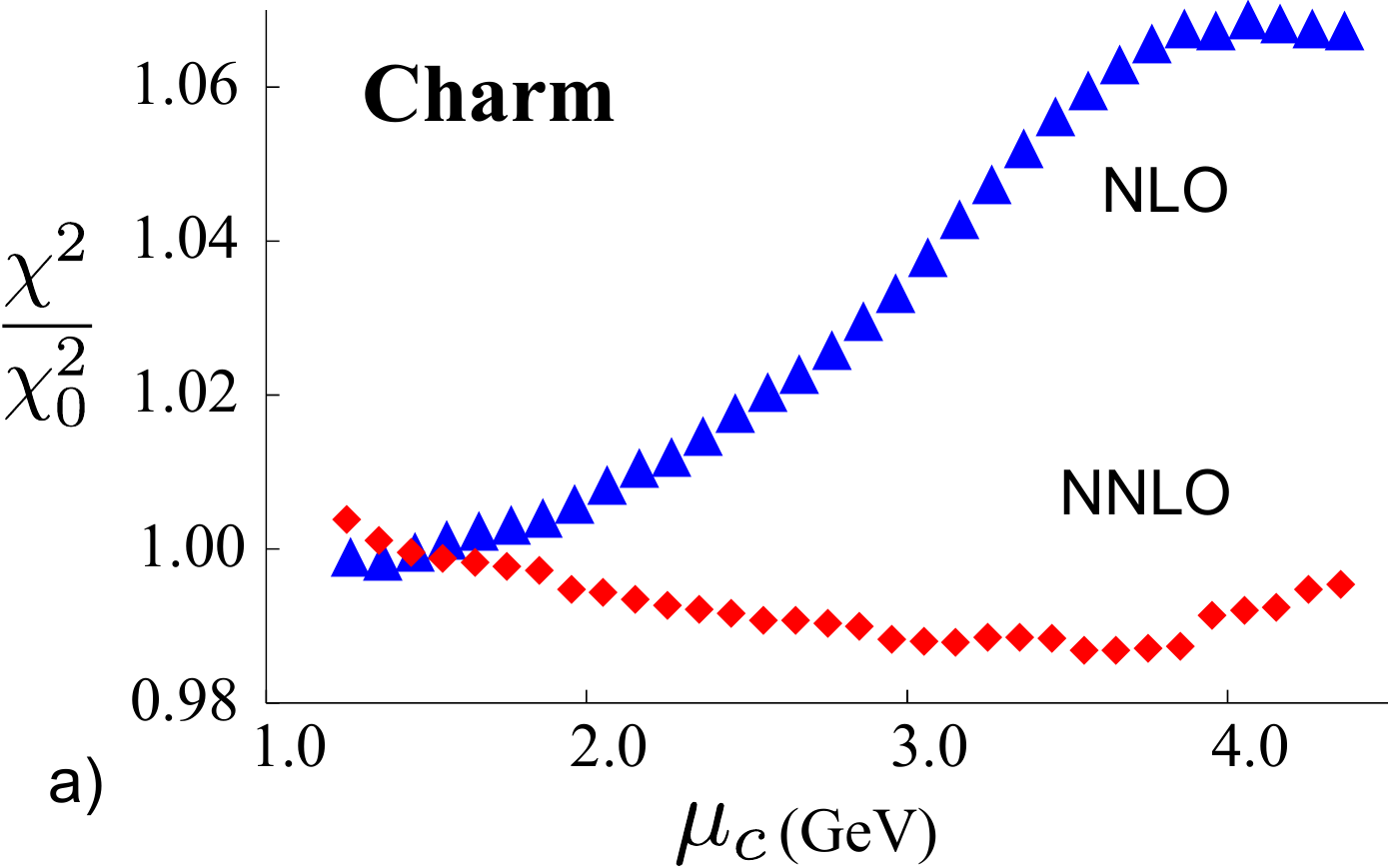}
\hfil
\includegraphics[width=0.45\textwidth]{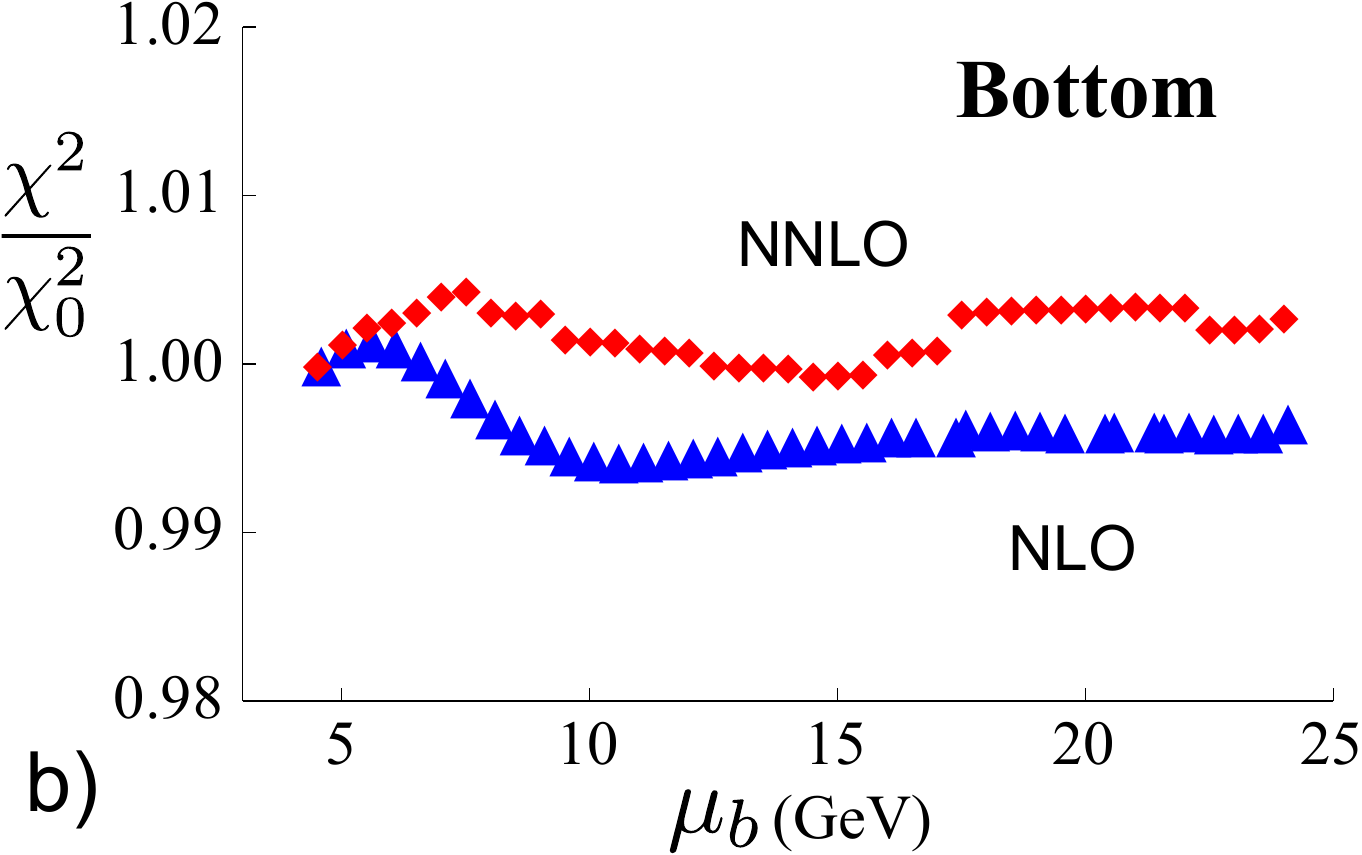}
\caption{
The ratio ($\chi^2/\chi^2_0$) of total $\chi^2$ values
(all data sets combined) 
as a function of the a)~charm and b)~bottom  matching scale $\mu_{c,b}$ in GeV. 
$\chi_{0}^{2}$ is the $\chi^{2}$ value for $\mu_{m}$ equal to the quark mass. 
The triangles (blue  \textcolor{blue}{$\blacktriangle$} )  are NLO
and the diamonds (red \textcolor{red}{$\blacklozenge$}) are NNLO. 
The fits are from Ref.~\cite{Bertone:2017ehk}.
\label{fig:chi2scaledi}
}
\vspace{-0.5cm}
\end{figure*}
}


\figone

\section{Introduction}
\figfrog

\vspace{-0.3cm}

The Parton Distribution Functions (PDFs) are the essential components
that allow us to make theoretical predictions for experimental
measurements of protons and hadrons.
The precision of the PDF analysis has advanced tremendously  in recent years, 
and these studies are now performed with very high  precision  at NLO and  NNLO in perturbation theory.
The xFitter project\footnote{%
xFitter can be downloaded from {\tt www.xFitter.org}.  An overview of
the program can be found in Ref.~\cite{Alekhin:2014irh}.  
} 
is an open source QCD fit framework 
that can perform PDF fits, assess the impact of new data, compare existing PDF sets, 
and perform a variety of other tasks~\cite{Alekhin:2014irh}.
The modular structure of xFitter allows for interfaces to 
a variety of external programs including: 
 QCDNUM~\cite{Botje:2010ay},
 APFEL~\cite{Bertone:2013vaa},
 LHAPDF~\cite{Buckley:2014ana},
 APPLGRID~\cite{Carli:2010rw},
 APFELGRID~\cite{Bertone:2016lga},
 FastNLO~\cite{fastnlo}  
and
 HATHOR~\cite{Aliev:2010zk}. 
A schematic of the modular structure is illustrated in Fig.~\ref{fig:flow}.

An overview of the recent xFitter updates and available tutorials 
is available in Ref.~\cite{Bertone:2017tig}. 
In this short report we will focus on the implementation of a 
generalized heavy quark matching scale $\mu_m$ and the implications for PDF 
fits.\footnote{A more extensive report of these features can be found in Ref.~\cite{Bertone:2017ehk}.}

\null \vspace{-0.8cm}
\section{The VFNS and FFNS}
\figvfnsi{}
\vspace{-0.3cm}

The inclusion of heavy quarks $Q=\{c,b,...\}$ into the PDF framework has been 
a formidable challenge. 
In the Fixed Flavor Number Scheme (FFNS), the heavy quark is excluded from the 
PDF parton-model framework; here, the heavy quark $Q$ must be produced explicitly 
such as in the process $\gamma g \to Q \bar{Q}$. 
In contrast, in the Variable Flavor Number Scheme (VFNS), the heavy quark is included
as a parton in the PDF at scales above the $\mu_m$ matching scale;\footnote{Details on the 
distinction between the matching and transition scales can be found in Ref.~\cite{Kusina:2013slm}
}
thus, 
we have the option of exciting a heavy quark $Q$ from within the proton, 
{\it e.g.} $\gamma Q \to Q g$.

Both the FFNS  and VFNS, as traditionally implemented, have advantages and disadvantages. 
The FFNS has the simplicity of avoiding an $N_F$ flavor threshold in the PDFs, 
but at large energy scales (such as at the LHC) the heavy quarks  $Q=\{c,b,...\}$ are treated 
differently from the light quarks.

Conversely, the VFNS has the advantage that it 
 resums the heavy quark contributions using the DGLAP evolution and treats 
all the quarks on an equal footing at large energy scales;
however, the  VFNS  can have some delicate cancellations when the heavy quark matching scale $\mu_m$
is similar to the heavy quark mass $m_H$.
Traditionally in most implementations of the VFNS, the 
heavy quark matching scale was chosen equal to the  heavy quark mass $\mu_m=m_H$
for a number of reasons as outlined in Ref.~\cite{Bertone:2017ehk,Kusina:2013slm}.
The new xFitter 2.0.0 program does not impose $\mu_m=m_H$, 
and  has the flexibility to choose any value for the matching scale $\mu_m$; 
thus, the difficulties of the traditional VFNS implementation with  $\mu_m=m_H$ are avoided. 
In a general sense, the variable matching scale allows us to interpolate continuously 
between the traditional VFNS (with $\mu_m=m_H$) and the FFNS (with $\mu_m\to \infty$).

This situation is summarized diagrammatically in Fig.~\ref{fig:vfns}.
In  Fig.~\ref{fig:vfns}-a), we see the traditional choice 
where the matching scale $\mu_m$ is set equal to the heavy quark mass $m_H$.
In  Fig.~\ref{fig:vfns}-b), we remove the  $\mu_m=m_H$ constraint and allow 
$\mu_m$ to take an arbitrary values. This is the new flexibility provided by xFitter 2.0.0.

\null \vspace{-0.9cm}
\section{Boundary Conditions}
\figBmatch{}
 \vspace{-0.3cm}

One of the key steps for implementing the variable heavy quark matching scales
is the correct boundary conditions between the $N_F$ and $N_F+1$ active flavors. 
These boundary conditions are displayed in Fig.~\ref{fig:bMatch}  for 
the case of the bottom quark PDF.

At NLO, if we match exactly at the bottom quark mass  $\mu_b=m_b$,
we find\footnote{This accidental 
cancellation for $\overline{MS}$  at NLO  was, in part, the reason for the 
traditional VFNS choice  $\mu_m=m_H$.
}
$f_b(x,\mu=m_b)=0$. 
For values  $\mu_b \not= m_b$, the boundary conditions are determined by the NLO contributions
from the DGLAP evolution kernels, and this is displayed in Fig.~\ref{fig:bMatch}-a). 
These contributions are driven by the $\ln(\mu/m_b)$ terms which are negative for $\mu<m_b$.
At  large $\mu$ scales, we observe the differences due to the choice of different 
boundary  conditions; this is due to the (un-resummed) higher order ${\cal O}(\alpha_S^2)$ terms 
which are not included at NLO.

In Fig.~\ref{fig:bMatch}-b) we display the NNLO matching conditions. 
In this case we find  \hbox{$x f_b(x,\mu) \not=0$} for  $\mu_b=m_b$.
At this order, we have included terms of one higher order in $\alpha_S$ compared to the previous case, 
and we see this tremendously reduces the variation of   $x f_b(x,\mu)$ for different 
choices of the matching scale $\mu_m$. 
This behavior is crucial as the choice of the heavy quark matching scale 
amounts to a scheme choice, and the resulting physics observables 
should be insensitive up to the corresponding order of perturbation theory.

\null \vspace{-0.9cm}
\section{Scheme Independence}
 \vspace{-0.3cm}

We can further illustrate the insensitivity of the physical observables to 
the choice of the heavy quark matching scale $\mu_m$ by examining the 
structure function $F_2^b(x,Q)$ displayed in Fig.~\ref{fig:fbottom}.
In Fig.~\ref{fig:fbottom}-a) we compute $F_2^b(x,Q)$ at NLO for 
a choice of $\mu_m$ values; at large energy scales $Q\sim 32\, $GeV we observe
a large dependence on the choice of $\mu_m$. 
In contrast, at NNLO in Fig.~\ref{fig:fbottom}-b) the 
variation of  $F_2^b(x,Q)$ is significantly reduced. 
Thus, the inclusion of the ${\cal O}(\alpha_S^2)$ NNLO contributions 
yields a result for the physical  $F_2^b(x,Q)$ which is very stable w.r.t. $\mu_m$.

Therefore, the NNLO implementation of the heavy quark matching scale 
in xFitter 2.0.0 has
eliminated many of the difficulties previously encountered with the 
NLO VFNS with the traditional choice of $\mu_m=m_H$.
    
\figFbottom{}

\null \vspace{-0.9cm}
\section{Impact on Fits}
 \vspace{-0.3cm}

To facilitate comparisons of the NLO and NNLO results,
Fig.~\ref{fig:chi2scaledi} displays the ratio $\chi^{2}/\chi^{2}_0$
for charm (on the left) and bottom (on the right) where $\chi^{2}_0$
is the value of the $\chi^2$ at $\mu_m=m_H$.
By plotting $\chi^{2}/\chi^{2}_0$, we can better compare the
fractional variation of $\chi^2$ across the matching scale values.\footnote{See
Ref.~\cite{Bertone:2017ehk} for the full details of the fit.}
At NLO for the case of charm, the optimal heavy quark matching scale
  for $\mu_c$ is in the general range $\mu_c\sim m_c$.
  For lower scales ($\mu_c \ll m_c$), $\alpha_S(\mu)$ is large and the
  charm PDFs are negative.  For higher scales ($\mu_c \gg m_c$),
  $\chi^2/\chi^2_0$ increases.
At NNLO for the case of charm, the $\chi^2/\chi^2_0$ variation
  is greatly reduced ($\sim 2\%$), and there is minimal sensitivity to the  $\mu_c$ scale in this range. 
For the case of bottom, the  the  $\chi^2/\chi^2_0$ variation
 is  very mild ($\sim 1\%$) for {\it both} NLO and NNLO; hence, the physics results are 
relatively insensitive to the particular choice of the heavy quark  matching scale $\mu_b$.

While the detailed characteristics of the above fits will depend on specifics 
of the analysis, there are two general patterns which emerge:
i)~the  $\chi^2$ variation of the NNLO results are generally reduced compared to the NLO results,
and
ii)~the relative $\chi^2$ variation across  the bottom transition is reduced compared to the charm transition. 
For example, although the global $\chi^2$ can be modified by different choices of data sets and
weight factors, these general properties persist across separate data sets.\cite{Bertone:2017ehk}
Additionally, there are a variety of prescriptions for computing the heavy flavor contributions;
these primarily differ in how the higher order contributions are organized. 
As a cross check, we performed a NLO fit using the  FONNL-A scheme; while
the absolute value of  $\chi^2$ differed, the above general properties persisted.

The net result is that we can now quantify the theoretical uncertainty
associated with the transition between different $N_F$ sub-schemes.
In practical applications, if we choose $\mu_{c}\sim m_c$, the impact
of the $N_F=3$ to $N_F=4$ transition is reduced as this is often below
the minimum kinematic cuts of the analysis ({\it e.g.} $Q_{min}^2$ and
$W_{min}^2$).
Conversely, the $N_F=4$ to $N_F=5$ transition is more likely to fall
in the region of fitted data; hence,  it is useful to quantify the
uncertainty associated with the $\mu_b$ choice.

\figChiScaledi{}

\null \vspace{-0.9cm}
\section{Conclusion}
 \vspace{-0.3cm}

The xFitter 2.0.0 program is a versatile, flexible, modular, and
comprehensive tool that can facilitate analyses of the experimental
data and theoretical calculations.
In this study we have examined the impact of the heavy flavor matching
scales $\mu_m$ on a PDF fit to the combined HERA data set.
These observations can be useful when performing fits. While charm has
a larger $\chi^{2}$ variation (especially at NLO), the charm quark
mass $m_c\sim 1.45$~GeV lies in a region which is generally excluded
by cuts in $Q^2$ and/or $W^2$.
On the contrary, the $\chi^{2}$ variation for the bottom quark is
relatively small at both NLO and NNLO. Since the bottom quark mass
$m_b\sim 4.5$~GeV is in a region where there is abundance of precision
HERA data, this flexibility allows us to shift the heavy flavor
threshold (and the requisite discontinuities) away from any particular
data set. Functionally, this means that we can analyze the HERA data
using an $N_F=4$ flavor scheme up to relatively large $\mu$ scales,
and then perform the appropriate NNLO matching (with the associated
constants and log terms) so that we can analyze the high-scale LHC
data in the $N_F=5$ or even $N_F=6$ scheme.

These variable heavy flavor matching scales $\mu_m$ allow us to
generalize the transition between a FFNS and a VFNS, and provides a
theoretical ``laboratory'' which can quantitatively test proposed
implementations.
In conclusion, we find that the ability to vary the heavy flavor
matching scales $\mu_m$, not only provides new insights into the
intricacies of QCD, but also has practical advantages for PDF fits.

\null \vspace{-0.9cm}
\begin{spacing}{1.0}

\end{spacing}

\end{document}